\documentclass[sigconf]{acmart}
\acmConference[ICSE 2024]{46th International Conference on Software Engineering}{April 2024}{Lisbon, Portugal}

\pdfoutput=1
\usepackage[english]{babel}

\usepackage{graphicx}
\usepackage{microtype}
\usepackage{booktabs}

\usepackage{float}
\usepackage{wrapfig}
\usepackage[normalem]{ulem}

\usepackage{bm}
\usepackage{textcomp}
\usepackage{amsmath}
\usepackage{amsthm}
\usepackage{amsfonts}
\usepackage{mathtools}
\usepackage{pifont}

\usepackage{algorithm}
\usepackage{algorithmic}

\usepackage{times}
\usepackage{gensymb}

\usepackage{textcomp}
\usepackage{multirow}
\usepackage{lscape}
\usepackage{subcaption}

\usepackage{mdframed}
\usepackage{hyperref}
\usepackage{textgreek}
\usepackage{tabularx, colortbl, xcolor}
\definecolor{mygray}{gray}{0.95}
\definecolor{mycyan}{HTML}{005397}
\definecolor{myred}{HTML}{E13333}
\definecolor{mymagenta}{HTML}{BF3E87}
\definecolor{mypurple}{HTML}{1B2278}

\definecolor{tearose}{HTML}{F584C5}
\definecolor{coral}{HTML}{F67088}
\definecolor{dodger_blue}{HTML}{3BA3EC}
\definecolor{domino}{HTML}{BC9F48}
\definecolor{domino}{HTML}{BC9F48}
\definecolor{domino}{HTML}{BC9F48}
\definecolor{catalina_blue}{HTML}{1C3168}
\definecolor{catalina_blue}{HTML}{1C3168}
\definecolor{catalina_blue}{HTML}{1C3168}
\definecolor{dark_scarlet}{HTML}{C63D52}
\definecolor{cerulean}{HTML}{0192A8}

\definecolor{tussock}{HTML}{C99E31}
\definecolor{p13}{HTML}{BFB5D7}
\definecolor{b14}{HTML}{BEA1A5}
\definecolor{y15}{HTML}{F0Cf61}
\definecolor{Merino}{HTML}{F3EEE3}

\newcolumntype{a}{>{\columncolor{p13}}l}

\usepackage[capitalize,noabbrev]{cleveref}
\crefname{ineq}{Inequality}{Inequalities}
\creflabelformat{ineq}{#2{\upshape(#1)}#3}

\theoremstyle{remark}

\newtheoremstyle{researchquestionstyle}{}{}{}{}{\bf}{}{.5em}{}
\theoremstyle{researchquestionstyle}

\usepackage{braket}

\usepackage{stmaryrd}

\usepackage{tikz}
\usetikzlibrary{trees}
\usepackage{tikz-dependency}
\usepackage{tikz-3dplot}
\usetikzlibrary{chains,scopes}
\usetikzlibrary{arrows.meta,automata,positioning}
\usetikzlibrary{shapes.geometric, arrows}
\usepackage{pgfplots}

\pgfplotsset{
  every axis/.append style = {thick},
  tick style = {thick,black},
  /tikz/normal shift/.code 2 args = {%
    \pgftransformshift{%
        \pgfpointscale{#2}{\pgfplotspointouternormalvectorofticklabelaxis{#1}}%
    }%
  },%
  shift/.style = {
    tick align        = outside,
    scaled ticks      = false,
    enlargelimits     = false,
    ticklabel shift   = {#1},
    axis lines*       = left,
    xtick style       = {normal shift={x}{#1}},
    ytick style       = {normal shift={y}{#1}},
    x axis line style = {normal shift={x}{#1}},
    y axis line style = {normal shift={y}{#1}},
  },
  shift/.default = 10pt,
  shift3d/.style = {
    shift=#1,
    ztick style       = {normal shift={z}{#1}},
    z axis line style = {normal shift={z}{#1}},
  },
  shift3d/.default = 10pt,
}

\usepackage{listings}
\usepackage{array}
\newcolumntype{H}{>{\setbox0=\hbox\bgroup}c<{\egroup}@{}}

\title{
Towards Precise Observations of Neural Model Robustness in Classification
}
\author{Wenchuan Mu}
\affiliation{
  \institution{Singapore University of Technology and Design}
  \country{Singapore}}
\email{wenchuan\_mu@sutd.edu.sg}

\author{Kwan Hui Lim}
\affiliation{
  \institution{Singapore University of Technology and Design}
  \country{Singapore}}
\email{kwanhui\_lim@sutd.edu.sg}

\copyrightyear{2024}
\acmYear{2024}
\setcopyright{rightsretained}
\acmConference[ICSE-Companion '24]{2024 IEEE/ACM 46th International Conference on Software Engineering: Companion Proceedings}{April 14--20, 2024}{Lisbon, Portugal}
\acmBooktitle{2024 IEEE/ACM 46th International Conference on Software Engineering: Companion Proceedings (ICSE-Companion '24), April 14--20, 2024, Lisbon, Portugal}\acmDOI{10.1145/3639478.3643519}
\acmISBN{979-8-4007-0502-1/24/04}
\begin{document}

\begin{abstract}

In deep learning applications, robustness measures the ability of neural models that handle slight changes in input data, which could lead to potential safety hazards, especially in safety-critical applications. Pre-deployment assessment of model robustness is essential, but existing methods often suffer from either high costs or imprecise results. To enhance safety in real-world scenarios, metrics that effectively capture the model's robustness are needed. To address this issue, we compare the rigour and usage conditions of various assessment methods based on different definitions. Then, we propose a straightforward and practical metric utilizing hypothesis testing for probabilistic robustness and have integrated it into the TorchAttacks library. Through a comparative analysis of diverse robustness assessment methods, our approach contributes to a deeper understanding of model robustness in safety-critical applications.
\end{abstract}

\maketitle

\section{Introduction}
\label{sec:intro}
Deep learning has attained significant accomplishments across a broad range of applications, including in systems critical to security such as self-driving cars, medical diagnosis, and face-recognition-based authentication systems. The reliability and robustness of deep neural networks (DNNs) are important in security-critical systems and for ensuring fair outcomes~\cite{halder2023capacity}. In such scenarios, even slight changes in input data can lead to catastrophic consequences, necessitating the pre-deployment assessment of model robustness.

The evaluation of model robustness is a well-established concept, but it comes with significant challenges. Existing robustness evaluation methods like adversarial testing and verification have their limitations. Adversarial testing may not accurately represent real-world scenarios, while verification often faces the issue of incomplete problem formulation~\cite{muller2022certified}. This means that verification methods might not fully capture the diversity of perturbations present in real-world scenarios. Furthermore, these methods may also encounter the problem of high cost, making them impractical for large-scale and resource-intensive applications~\cite{zhang2023proa}. Hence, there is a need for broader, practical evaluation methods for robustness assessment.

In order to address these gaps and bolster the safety of deep learning applications, our research focuses on the probabilistic robustness assessment. While some existing probabilistic robustness evaluations resort to approximated methods, these approximations may lead to the omission of critical adversarial instances, consequently overestimating the true robustness of the model.

In our work, we integrate the exact binomial test into the robustness evaluation of deep neural networks (DNNs), implemented within the TorchAttacks library (available at \url{https://github.com/cestwc/precise-robustness}). The exact binomial test is a statistical method that precisely measures how small changes in inputs affect the output of DNNs. This technique provides a clear and accurate way to identify vulnerabilities in neural models. Our method is notable for its efficiency, requiring less computational resources compared to traditional methods. It is versatile and can be applied to various DNN architectures, making it a practical solution for assessing robustness in safety-critical applications.

\section{Probabilistic Robustness From Binomial Testing}

There exist multiple interpretations of classifier robustness and we opt for the definition that emphasises the probabilistic nature of adversarial examples. Formally, $P_\mathbf{x} (P(h(\bm{x'}) \neq h(\bm{x})\mid \mathbf{x}=\bm{x}, d(\bm{x}, \bm{x'}) \le \epsilon) \leq \kappa)$, where $\mathbf{x}$ is the random variable input in the distribution, $\bm{x}, \bm{x'}$ are specific inputs, $d$ denotes distance, $\epsilon$ denotes an imperceptible perturbation. To calculate the probability of any sampled input has less than $\kappa$ (\emph{e.g.}, 1\%) adversarial examples in its neighbourhood, we first formulate this event as a Bernoulli trial $\mathbf{z}$, where the true probability is $P_\mathbf{x}(\mathbf{z}=1\mid h)$.

\paragraph{Binomial Test With Exact Solution}

To get $P_\mathbf{x}(\mathbf{z}=1\mid h)$, we may first address $P(\mathbf{z}=1\mid h, \mathbf{x}=\bm{x})$ at specific $\bm{x}$. Given $\bm{x}$, we want to determine if the probability that $h$ makes an incorrect prediction around $\bm{x}$ is greater than or equal to $\kappa$. This forms the null hypothesis in an exact binomial test. In a right-tail exact binomial test,
\begin{equation}
    P(\mathbf{w} = 0 \mid h, \mathbf{x}=\bm{x}) = \sum_{i=k}^{n} \binom{n}{i} p_{\bm{x}}^i (1-p_{\bm{x}})^{n-i}
\end{equation}
where $n$ denotes sample size, $k$ denotes the number of successes, $p_{\bm{x}}$ is the true probability of success, and $\mathbf{w}$ is the observed event that the total number of successes is less than $k$. For the given $\bm{x}$, we increase the sample size until it rejects either side of the tail, \emph{i.e.}, we would have high confidence ($1-\alpha$) to know that 
\begin{equation}
    P(\mathbf{w} = 1 \mid p_{\bm{x}} > \kappa) < \alpha
\end{equation} and similarly on the other tail we have
\begin{equation}
    P(\mathbf{w} = 0 \mid p_{\bm{x}} < \kappa) <\alpha.
\end{equation}

\paragraph{True Probability Rather Than Observed Events}

Existing works also stop at rejecting the null hypothesis and calculate the frequency of right rejection. However, we claim that the frequency of right rejection, \emph{i.e.}, the probability of observation of event $\mathbf{w}$, or $\mathbf{w} = 1$, is not the true probability we are looking for.

Instead, we shall always compute the probability that event $\mathbf{z}$ is true. To achieve that, we 
apply the law of total probability. Then, the probability of event $\mathbf{w}$ can be expressed as
\begin{equation}
    P(\mathbf{w}) = P(\mathbf{w}\mid \mathbf{z})P(\mathbf{z}) + P(\mathbf{w}\mid \lnot \mathbf{z})P(\lnot \mathbf{z})
\end{equation}
If we further write $P(\lnot \mathbf{z}) = 1 - P(\mathbf{z})$, $P(\mathbf{w}\mid \lnot \mathbf{z}) = 1 - P(\lnot \mathbf{w}\mid \lnot \mathbf{z})$, we eventually get
\begin{equation}
    P(\mathbf{z}) = \frac{P(\mathbf{w}) - P(\mathbf{w}\mid \lnot \mathbf{z})}{1 - P(\lnot \mathbf{w}\mid \mathbf{z}) + P(\mathbf{w}\mid  \lnot \mathbf{z})}
\end{equation}
Now that we know that $0 < P(\mathbf{w}\mid \lnot \mathbf{z}), P(\lnot \mathbf{w}\mid \mathbf{z}) < \alpha$, we can find the lower and upper limit of $P(\mathbf{z})$ as 
\begin{equation}
    (P(\mathbf{w}) - \alpha)/(1+\alpha) < P(\mathbf{z}) < P(\mathbf{w}) / (1 - \alpha)
\end{equation}
which makes sense because $P(\mathbf{z})$ is still predominantly positively related to $P(\mathbf{w})$, while the smaller false positive rate ($\alpha$) we have the closer $P(\mathbf{z})$ will be to $P(\mathbf{w})$.

In this way, we have made our observation targeted on the true probability, instead of the samples. Conservatively, a simple way is to get the $P(\mathbf{w})$ first, subtract the false positive rate from it, and divide by (1 + false positive rate).

To complete the process, we still need to determine $P(\mathbf{w})$. If in $n'$ times we observed $\mathbf{w}$ $k'$ times and not $\mathbf{w}$ $n'-k'$ times, then we can calculate the probability of $\mathbf{w}$ given these observations using the likelihood $P(\mathbf{w}) = k'/n'$. In summary, we get
\begin{equation}
    \frac{k'/n' - \alpha}{1+\alpha} < P_\mathbf{x}(\mathbf{z}=1\mid h) < \frac{k'}{n'(1 - \alpha)}    
\end{equation}

\section{EXPERIMENTS}
\begin{table}
\centering
    \caption{Classification results on CIFAR-10. Our observations of robustness and popular attack failure rates are listed side by side. Our observation gives the minimum probability that the adversarial examples of an arbitrary input account for less than 1 in 10,000.}
    \begin{tabular}{l*{3}{|>{\centering\arraybackslash}p{0.18\linewidth}}}
\toprule
Training & Accuracy   & Attack Failure Rate & Our Observation  \\
\midrule
ERM~\cite{vapnik1999nature}      & \textbf{94.38} & 1.25          & 84.20          \\
ERM+DA\cite{shorten2019survey}   & 94.21 & 1.08          & 84.15         \\
FGSM~\cite{goodfellow2014explaining}     & 84.96 & 43.50          & 83.50          \\
PGD~\cite{madry2017towards}      & 84.38 & 47.07         & 82.90          \\
TRADES~\cite{zhang2019theoretically}   & 80.42 & \textbf{48.54}         & 79.12         \\
MART~\cite{wang2019improving}     & 81.54 & 48.90          & 80.21         \\
PRL~\cite{robey2022probabilistically}      & 93.82 & 0.71          & \textbf{90.63}         \\
\bottomrule
\end{tabular}
    \label{tab:acc}
\end{table}

We conduct experiments on the CIFAR-10 dataset. We estimate 6 popular robustness improvement models from ERM~\cite{vapnik1999nature}: ERM+DA~\cite{shorten2019survey}, FGSM~\cite{goodfellow2014explaining}, PGDT~\cite{madry2017towards}, TRADES~\cite{zhang2019theoretically}, MART~\cite{wang2019improving}, and PRL~\cite{robey2022probabilistically}.
We also compare our robustness estimation (lower bound) with vanilla accuracy and attack-failure rate using projected gradient descent~\cite{madry2017towards}.

We use our method to evaluate existing adversarial mitigation methods on the CIFAR-10 dataset, with the result presented in \cref{tab:acc}. ERM leads in accuracy with 94.38\%, while MART, known for its state-of-the-art adversarial training, records the highest Attack-Failure Rate at 48.59\%. In contrast, the PRL method excels in robustness estimation, achieving a significant score of 90.63\%. This performance underscores PRL's capability to improve probabilistic robustness (a critical attribute for models in safety-critical applications). It is important to note the distinct focus of each model: ERM prioritizes accuracy without significant emphasis on robustness, MART leverages adversarial attacks for robustness training, and PRL employs probabilistic methods for robustness training. The respective best performances in their focused areas validate the strengths of our approach, particularly highlighting the balance between robustness and accuracy estimation achieved by our method, which is vital in contexts where neither high accuracy nor attack resistance alone suffices.

\paragraph{Conclusion}

This study introduces a new method to improve the assessment of probabilistic robustness in neural networks against adversarial examples, comprising three main elements: an exact binomial test for accurate binomial distribution calculations, a technique to reduce degrees of freedom based on the law of total probability, and standardized failure rate thresholds. Our exact solution addresses potential certification errors caused by approximations. The approach aligns better with the concept of probabilistic robustness by reducing unnecessary false positive rates, using IEC 61508 for certification thresholds to match safety integrity levels.

\vspace{3mm}
{
{\noindent Acknowledgments.} 
This research is supported by the Ministry of Education, Singapore, under its Academic Research Fund Tier 2 (Award No. MOE-T2EP20123-0015). Any opinions, findings and conclusions, or recommendations expressed in this material are those of the authors and do not reflect the views of the Ministry of Education, Singapore.
}

\end{document}